\documentclass[aps,prl,twocolumn,groupedaddress]{revtex4}
\usepackage{graphicx}% Include figure files

\begin{document}

\title{Ghosts in a Mirror}

\author{Alexander Feinstein and Sanjay Jhingan}

\affiliation {Dpto. de F\'{\i}sica Te\'orica, Universidad del
Pa\'{\i}s Vasco, Apdo. 644, 48080, Bilbao, Spain}

\date{\today}

\begin{abstract}
  We look at some dynamic geometries produced by scalar fields
  with both the ``right" and the ``wrong" sign of the kinetic
  energy.  We start with anisotropic homogeneous universes with
  closed, open and flat spatial sections. A non-singular solution
  to the Einstein field equations representing an open anisotropic
  universe with the ghost field is found. This universe starts
  collapsing from $t \to -\infty$ and then expands to $t \to
  \infty$ without encountering singularities on its way. We
  further generalize these solutions to those describing
  inhomogeneous evolution of the ghost fields. Some interesting
  solutions with the plane symmetry are discussed. These have a
  property that the same line element solves the Einstein field
  equations in two mirror regions $\left|t\right|\geq z$ and
  $\left|t\right|\leq z$, but in one region the solution has the
  \emph{right} and in the other, the \emph{wrong} signs of the
  kinetic energy. We argue, however, that a physical observer can
  not reach the mirror region in a finite proper time.
  Self-similar collapse/expansion of these fields are also briefly
  discussed.
\end{abstract}


\maketitle

\section{Introduction}\label{intro}
Recently some authors have discussed scalar fields with negative
kinetic energies (NKE) \cite{Carroll,Gibbons,Nojiri}. In
\cite{Carroll} the authors have studied these fields in connection
to the dark matter problem in the universe. Also, since
observational evidence does not exclude cosmological models with
the pressure to density ratio $<-1$, various models with negative
densities were studied recently by several groups \cite{negative}.

In \cite{Gibbons}, to motivate these studies, it was argued that
one may find several physical examples, such as Lifshitz
transitions in condensed matter physics or an unusual dispersion
relation for rotons in liquid helium, suggesting that the NKE may
appear in nature. Also, negative energy densities may appear in
quantum particle creation processes in curved backgrounds
\cite{Candelas} or in squeezed states of the electromagnetic
fields \cite{Slusher}. Their appearance, however, signals usually
the unhealthy vacuum instability. Therefore, if the time scale of
this instability is too short, the matter cannot serve as viable
candidate for the dark energy component \cite{Carroll}. A
different reasoning towards NKE, however, may be applied if one
approaches the cosmological singularity problem.

One believes that the emergence of spacetime singularities in
General Relativity suggests the breakdown of the theory at its
natural scales. At these scales one expects the quantum
corrections to take a leading role and save the situation.
Different approaches to regularize the singularity have been
undertaken. Phenomenologically, and in the light of all the
standard singularity theorems, what appears to be the simplest way
to tackle the singularity problem is to allow for negative energy
densities, and probably some degree of anisotropy and/or
inhomogeneity. This would produce the repulsive gravity effects
and may smoothen the singularity.  Phenomenologically, again, one
can associate the negative energy densities with the back reaction
of the quantum fluctuations, so that the idea in itself is not
that ridiculous. Thus, allowing for NKE may prove to be an
interesting approach to spacetime singularities in General
Relativity. Here, the problem of the vacuum instability should be
irrelevant, and even may become a blessing, for if these exotic
fields decay rapidly, after having smoothed the singularity, this
well could be a reason as to why we live in a ghost-free world.
Consequently, it is worthwhile to have a closer look at some
geometries produced by such fields, and we suggest here that one
should keep her/his eyes wide open, just in case.

We will be especially interested in dynamical spacetimes in this
setting, some static solutions in the ghost sector were presented
elsewhere \cite{Gibbons,Gibbons-Rasheed}.  Just to concentrate on
the simplest examples we consider the energy-momentum tensor of a
massless scalar field in the following form,
\begin{equation}
  \label{eq:emtensor}
  T_{\mu \nu} = \epsilon(2 \varphi_{,\mu} \varphi_{,\nu} - g_{\mu\nu}
  \varphi_{,\alpha}\varphi^{,\alpha}) ,
\end{equation}
which is derived from the Lagrangian
\begin{equation}
  \label{eq:action}
{\cal L} = {\cal R} -2 \epsilon\varphi_{,\alpha}\varphi^{,\alpha} ,
\end{equation}
and where $\varphi$ is a scalar, ${\cal R}$ is the Ricci scalar
and $\epsilon$ may take values $1$ and $-1$.  Our metric signature
convention is $(-,+,+,+)$. When $\epsilon = 1$, the
energy-momentum tensor is of a standard form, while $\epsilon =
-1$ stands for the ghost field.

\section{HOMOGENEOUS ANISOTROPIC MODELS}\label{H-A-M}
Our starting point is the family of Kantowski-Sachs metrics. We
choose these space-times because they combine both the non-trivial
curvature as well as the effects of anisotropy. We should note,
moreover, that we couldn't find any other non-pathological,
dynamical homogeneous and isotropic spacetime with the ghost
fields.

With the energy-momentum tensor given by (\ref{eq:emtensor}) we
find the following formal solution to the Einstein field
equations,
\begin{eqnarray}\label{Kantowski-Sachs}
    ds^2=-(\frac{2\eta}{t}-k)^{-\alpha}dt^2+
    (\frac{2\eta}{t}-k)^{\alpha}dr^2  \nonumber \\
    + t^2(\frac{2\eta}{t}-k)^{1-\alpha}(d\theta^2+f_k(\theta)^2d\phi),
\end{eqnarray}
with the scalar field given by
\begin{equation}
  \label{eq:KS-sol}
  \varphi(t)=\frac{1}{2}\sqrt{\frac{1-\alpha^2}{\epsilon}}
  \log\left(\frac{2\eta}{t}-k\right).
\end{equation}
Here $k$ is spatial curvature $(k = -1,0,1)$, and
\begin{equation}
  \label{eq:ftheta}
  f_k(\theta)=\left\{
      \begin{array}{ll}
\sin \theta  & k=1 , \\
 \theta     & k=0 , \\
\sinh \theta &  k=-1 .
      \end{array}\right.
\end{equation}
Now, for $\epsilon=1$ and $|\alpha| < 1$, we recover the usual
Einstein-dilaton solution given, for example, in \cite{AF-VM}.

When $\epsilon=-1$ one must continue the scalar field analytically
in order to finish with a real field and a non-pathological
metric.  In this case one may obtain real solutions simply by
considering the case $|\alpha|>1$.  When $\alpha=0$, however, we
must continue analytically the $log$ function. It follows, thus,
that it is only possible to obtain real solutions in the
homogeneous case when $k\neq0$. On the other hand, when certain
amount of inhomogeneity is introduced the extension of $k=0$
solutions is also possible. We will discuss these cases in the next
section.

The two ghost solutions with $\alpha=0$  are:
\begin{itemize}
\item $k=1$
  \begin{eqnarray}
    \label{eq:Kone}
    ds^2& = &-dt^2+dr^2-(\eta^2+t^2)(d\theta^2+\sin^2\theta d\phi^2) ,\\
    & & \varphi(t)=\arctan\left(\frac{\eta}{t}\right) \nonumber,
  \end{eqnarray}
\item $k=-1$
  \begin{eqnarray}
    \label{eq:Kmnone}
    ds^2&=&-dt^2+dr^2+(\eta^2+t^2)(d\theta^2+\sinh^2\theta d\phi^2) ,\\
    & & \varphi(t) = \arctan\left(\frac{\eta}{t}\right) \nonumber.
  \end{eqnarray}

\end{itemize}
The solution (\ref{eq:Kone}) is just the Gibbons-Rasheed's
massless ``ghost'' wormhole \cite{Gibbons-Rasheed} with $t$ now
being the spacelike and $r$ the timelike co-ordinates. The
solution (\ref{eq:Kmnone}), however, represents a
\emph{non-singular} anisotropic universe with open spatial
sections. Its volume expansion $\Theta$ is :
\begin{equation}
  \label{eq:expansion}
  \Theta=\frac{2t}{\eta^2+t^2} .
\end{equation}
\begin{figure}
 \centering \includegraphics[width=7cm,height=4cm]{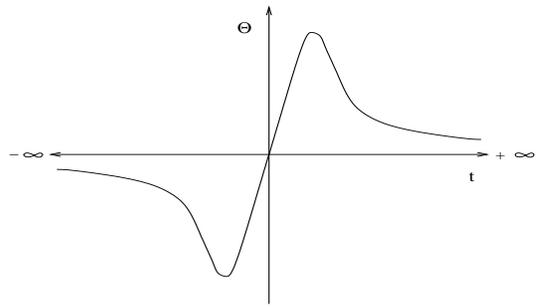}
  \caption{{Volume expansion of nonsingular universe}
  \label{fig:expansion}}
\end{figure}

The volume deformation reads:
\begin{equation}
  \label{eq:shear}
  \sigma= -\frac{t}{\sqrt{3}(\eta^2+t^2)}
\end{equation}
The average scale factor $a(t)=(\eta^2+t^2)^{1/6}$, behaves as
$a\propto t^{1/3}$ for $\left|t\right|>> \eta$, proper of a stiff
fluid.

The non-singular universe is flat and static as $t\to -\infty$,
with the vanishing scalar field. It collapses towards a flat
universe at $t=0$, where the scalar stabilizes to a constant, and
then expands again to a flat universe as $t\to \infty$, with the
vanishing scalar field.  The solution (\ref{eq:Kmnone}) is
definitely too crude to represent even a toy cosmology. But
imagine, that the massless scalar field with the NKE is
supplemented by an ordinary matter, or by a potential as in
\cite{Carroll}. Then, it may be possible that the evolution of the
model at different stages is dominated by either the ghost field
or by ordinary matter. Near the singularity, the kinetic terms
always dominate the potential, and therefore the short period of
the ghost domination may give rise to a regular solution. There
could also be situations where both ingredients will just conspire
to produce a cosmological constant, as in \cite{Gibbons}, during
some epoch. Such a model could then be an interesting test bench
to address the singularity, inflation and the coincidence problem
at one go. The realistic model building, however, is far beyond
the scope of the present paper.

In a different setting, one can make a connection between the
solution (\ref{eq:Kmnone}) and the singularity problem in low
energy string cosmology. The related models are those of the
pre-Big Bang scenario (for a review see, \cite{Lidsey,Gasperini}).
One of the main difficulties in these models is the smooth exit
from the pre-Big Bang into the post-Big Bang regime. The
possibilities of the smooth transitions in the context of the
lowest-order effective string action were discussed in
\cite{Risi}. There it was concluded that the transitional
singularities at $t=0$ can be avoided for anisotropic backgrounds
provided one accepts the sources with negative energy densities.
The solution (\ref{eq:Kmnone}) therefore represents an exact
evolution of such a model near $t=0$.

\section{Inhomogeneous solutions}\label{I-S}
When $k=0$, the spatially flat case, the two dimensional line
element $(d\theta^2 + \theta^2 d\phi^2)$ can be cast into an
explicitly flat form: $dx^2+dy^2$. In this case the analytical
continuation of the scalar field simply does not exist. We,
therefore, will allow for a certain degree of inhomogeneity and
consider the following form of the plane symmetric solutions,
\begin{equation}
  \label{eq:plane}
  ds^2=\frac{1}{\sqrt{t}} e^{f(t,z)} (-dt^2+dz^2)+t(dx^2+dy^2) .
\end{equation}
It should be noted that the discussion here can be generalized in
a straightforward manner to any $G_2$ spacetime with two commuting
spacelike Killing vectors by introducing transversal gravitational
degrees of freedom in the following form
$(e^{P(t,z)}dx^2+e^{-P(t,z)}dy^2)$. The off-diagonal terms may
also be included. For the sake of clarity, however, we stick to a
simple plane symmetric case, which we feel is sufficient to make
our point.

Assuming the geometry (\ref{eq:plane}) the Klein-Gordon equation
reads
\begin{equation}
  \label{eq:K-G}
  {\varphi_{,tt}}+\frac{1}{t}{\varphi_{,t}}-{\varphi_{,zz}}=0,
\end{equation}
and when $\varphi(t,z)$ is the solution of this equation, the metric
function $f(t,z)$ can be obtained by quadratures:
\begin{eqnarray}
  \label{eq:quadrature}
  f_{,t}&=&2 \epsilon t ({\varphi_{,t}}^2+{\varphi_{,z}}^2) \\
  f_{,z}&=& -4 \epsilon {\varphi_{,t}}{\varphi_{,z}} .
\end{eqnarray}
Again, if $\epsilon$ is $1$ we are dealing with the standard
scalar field, and when $\epsilon=-1$ it is a ghost. Thus, the NKE
solutions are related to the positive kinetic energy solutions by
$f\to -f$ transformation.

We consider now two possible solutions to (\ref{eq:K-G}):
\begin{equation}
  \label{eq:sol_zgt}
  \varphi_1= b \; \hbox{arccosh}\left(\frac{z}{t}\right), \qquad{|z|
  \geq t}
\end{equation}
and
\begin{equation}
  \label{eq:sol_zlt}
  \varphi_2= b \; \hbox{arccos}\left(\frac{z}{t}\right), \qquad{|z|
  \leq t} .
\end{equation}
Note that $\varphi_1$ relates to $\varphi_2$ by $b\to i b$
transformation even though they ``operate'' in two different,
mirror, regions of spacetime. The positive energy solutions of
Einstein equations for $\varphi_1$ and $\varphi_2$ are:
\begin{equation}
  \label{eq:phi_1}
  ds^2 [1]=\frac{1}{{\sqrt t}}
  (\frac{z^2}{t}-t)^{-2 b^2}(-dt^2+dz^2)+t(dx^2+dy^2),
\end{equation}
for $|z| \geq t$, and
\begin{equation}
  \label{eq:phi_2}
  ds^2[2]=\frac{1}{{\sqrt t}}
(t-\frac{z^2}{t})^{2 b^2}(-dt^2+dz^2)+t(dx^2+dy^2),
\end{equation}
for $ |z| \leq t$. It is amazing, however, that the positive and
the negative energy solutions interchange when the kinetic energy
sign flips over and the power $2 b^2$ is even: the negative energy
solution for $\varphi_1$, is given by the line element
(\ref{eq:phi_2}), while the negative energy solution for
$\varphi_2$ is given by (\ref{eq:phi_1}). Therefore, the same
geometry (\ref{eq:phi_1}) is induced by either the normal scalar
field in region $|z| \geq t$, or by a ghost field in its mirror
region $|z| \leq t$. The same applies to (\ref{eq:phi_2}).

Now, for the solution (\ref{eq:phi_1}) Ricci scalar is
\begin{equation}
  \label{eq:Ricci_1}
{\cal R} = -2 b^2 t^{-3/2}(\frac{z^2}{t}-t)^{2 b^2},
\end{equation}
and, for the solution (\ref{eq:phi_2}),
\begin{equation}
  \label{eq:Ricci_2}
{\cal R} = 2 b^2 t^{-3/2} (t-\frac{z^2}{t})^{-2 b^2}.
\end{equation}
Therefore, the solutions (\ref{eq:phi_2}) are singular at $t=z$,
but (\ref{eq:phi_1}) are not. Of course, there happen to be
singularities at $t=0$ in these expressions, but these are not of
our interest in this section. We have further checked the other
curvature scalars for (\ref{eq:phi_1}), such as square and the
cube of the Ricci and the Riemann tensors and found those regular
at $t=z$.  The possibility then arises, to continue the solution
(\ref{eq:phi_1}) as it stands into the region $|z|<t$, by flipping
the sign of the action. The spacetime could have looked as
depicted in fig. \ref{fig:mirror} below.
\begin{figure}[h]
  \centering \includegraphics[width=10cm,height=5cm]{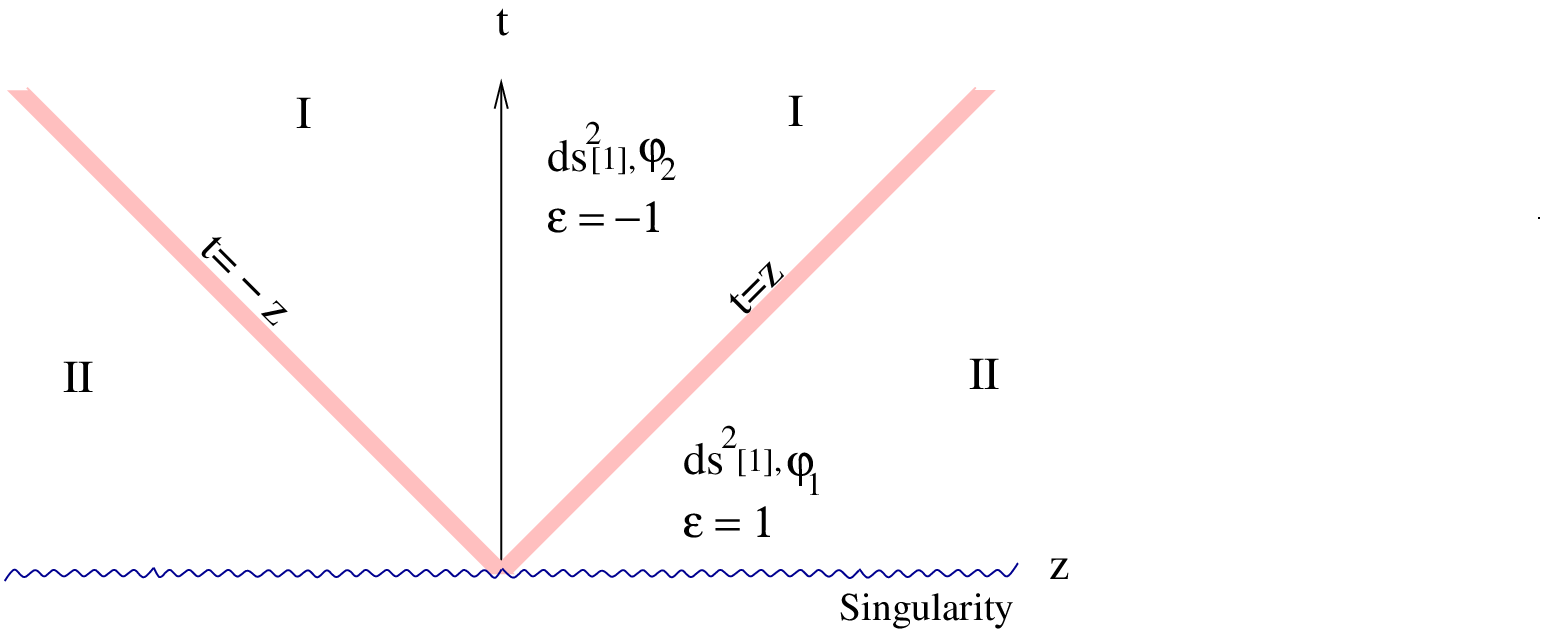}
  \caption{Mirror solutions}
  \label{fig:mirror}
\end{figure}

The picture looks quite unexpected. One has equivalent geometry on
both sides of the mirror line $z=t$, produced by the standard scalar
field in the region II, while the same geometry in the region I is
produced by a ghost.  Technically this happens because the pure
imaginary transformation of the scalar field $\varphi_1$ into $\varphi_2$
produces a real scalar field in a mirror region.  It is tempting to
call such scalar fields as mirror-ghosts. A different pair of scalar
fields which solve the equation (\ref{eq:K-G}) and have the mirror
ghost properties are $\varphi =b/\sqrt{t^2-z^2}$ and $\varphi
=b/\sqrt{z^2-t^2}$.

On a closer inspection of the geometry, described by the fig.
\ref{fig:mirror}, one finds that the co-ordinates $t$ and $z$ are
not that appropriate to describe the situation. It is convenient
to introduce the following chart $t=u^n +v^n$ and $z=u^n - v^n$,
where $n=1/(1-2b^{2})$ \cite{Feinstein}, and since we have
insisted that $2b^2$ is even, the power $n$ must be negative.  It
is then easy to see that the surface $z=t$ corresponds to $v^n=0$
and it cannot be reached by a physical observer in a finite proper
time.  Thus the two regions are \emph{worlds apart} for their
internal observers. The ``mirror" and the ``real" world do not mix
up!\footnote{The idea of
  mirror worlds was introduced many years ago by Yu. Kozarev, L. Okun
  and E.  Pomeranchuck \cite{Pomeranchuk}. Glashow \cite{Glashow} has
  shown that the particles of the two worlds cannot interact. It is
  interesting that we come to a similar conclusion and from a
  completely different direction.}

We now consider inhomogeneous evolution with nontrivial spatial
curvature. To do so, we generalize the $\alpha=0$ solutions of the
previous section using the following anzatz:

\begin{equation}\label{roberts}
    ds^2=-dudv + R^2(u,v) (d\theta^2+f_k(\theta)^2 d\phi^2) .
\end{equation}
with $f_{k}(\theta)$ as defined in Eqn. (\ref{eq:ftheta}).

With the energy-momentum tensor given by (\ref{eq:emtensor}) we find
the following solution:
\begin{eqnarray}\label{R-sol}
\varphi(u,v) &=& \frac{1}{2{\sqrt \epsilon}} \ln\left[\frac{(1-
{\sqrt \epsilon}p) kv - u}{(1+{\sqrt \epsilon}p)k v-u}\right] , \\
R^2(u,v) &=&\frac{1}{4}[(1-\epsilon p^2)k^2v^2-2kvu+u^2] ,
\end{eqnarray}
where the parameter $p$ represents the strength of the scalar
field. For $k=1$ and $\epsilon=1$ the solution reduces to the one
found by Roberts \cite{Roberts}. It was further thoroughly
investigated by various authors in connection with the spherically
symmetric scalar field collapse (see, for example, \cite{ONT}). In
the standard scalar field case, the $k=0$ solutions are flat
spacetimes written in plane wave co-ordinates. While, the $k=-1$
represents scalar field collapse/expansion with open geometry
where one does not expect black hole formations since the trapped
surfaces, if formed, are non-compact.  Thus, in the $k=-1$ case
one either has a naked singularity or a regular collapse/expansion
similar to the case of $k=1$ \cite{ONT}.

We now consider the phantom solutions ($\epsilon=-1$). These are given by
\begin{equation}
  \label{eq:phantom3}
  \varphi(u,v)= \hbox{arctan}\left(\frac{pkv}{kv-u}\right) .
\end{equation}
One can see that the scalar field remains regular everywhere. As a
consequence, the scalar curvature is also regular everywhere, but
in the geometrical centre $u=0$ and $v=0$:
\begin{equation}\label{cscalepmn1}
{\cal R} = -\frac{8k^2 p^2 u v}{[p^2k^2v^2+(kv-u)^2]^2}.
\end{equation}
The higher order curvature invariants behave in a similar way. It
is further informative to look at the character of the gradient of
the area coordinate $R(u,v)$, i.e.,
$R_{\alpha}(u,v)R^{\alpha}(u,v)$  \cite{Senovilla}, which is given
by
\begin{equation}\label{apparent}
R_{\alpha}R^{\alpha}=k-k^2\frac{p^2uv}{p^2k^2v^2+(kv-u)^2}
\end{equation}
The character of the gradient of the area coordinate tells one
about trappedness of the surfaces.  It follows, that for a given
$k$, the character of the area gradient never changes the sign
(this doesn't happen in the non-ghost case where the area gradient
is parameter $p$- sensitive). Note, that the change $k=1$ to
$k=-1$ corresponds to a change $v\to-v$, or, in $t$ and $r$
co-ordinates ($t=u+v$ and $r=u-v$) to an $r \leftrightarrow t$
interchange, exactly as in the homogeneous cases (\ref{eq:Kone}),
(\ref{eq:Kmnone}).  Therefore, the solutions with $k=1$ should be
seen as dynamical generalizations of the Gibbons-Rasheed worm hole
\cite{Gibbons-Rasheed}, and in fact if looked at carefully in
($t,r$) coordinates represent a time sequence of static worm
holes. The solutions with $k=-1$ are inhomogeneous generalizations
of the solution (\ref{eq:Kmnone}).  This interpretation is further
stressed by the behaviour of the area gradient, being globally
timelike (or null) in the cosmological case, and globally
spacelike (or null) in the case of $k=1$.

One can as well look at the scalar curvature singularity at $u=0$
$\cap$ $v=0$ in $t$ and $r$ coordinates.  The singularity occurs
at $R(t,r)=0$. In the $k=1$ case this surface is given by the
equation $ r^2 +p^2(t-r)^{2}/4 =0$, while in the $k=-1$ case, it
is $ t^2 +p^2(t-r)^{2}/4 =0$. Therefore, the singularity occurs at
$t=r=0$.

One may interpret the $k=1$ solution along the lines of
\cite{ONT}. These, then represent non-singular ghost field
collapse.  The ghost fields should be thought of as imploding from
the past null infinity, ($u\to -\infty$). One can easily check
that the scalar field vanishes on $v=0$ and is constant along
$u=0$ null hypersurfaces. The mass function vanishes on both and
so does the flux across these null hypersurfaces.  Therefore, one
can match $v<0$ and $u>0$ regions with flat spacetime \cite{ONT},
avoiding singularities. We note that the parameter $p$ plays no
essential role here.

\section{Conclusions}

We have looked at various geometries produced by massless scalar
fields both with the positive and the negative kinetic energies.
In the homogeneous case, we have shown the existence of
non-singular anisotropic solution with open spatial sections and
NKE. Assuming inhomogeneous plane symmetry we have found mirror
images with exactly the same geometry in different regions of
spacetime produced by the scalar fields with positive (in one
region) and the negative (in the other) kinetic energies. We have
shown, however, that the physical observers from the positive
energy region cannot reach the phantom geometry in a finite proper
time. We have further generalized the self-similar scalar field
dynamical solutions to include both the nontrivial curvature and
the negative NKE. These generalize the static ghost worm hole
solution to the dynamical case ($k=1$), and the homogeneous
anisotropic universe to the inhomogeneous one ($k=-1$). The
generalizations introduce singularities which can, probably, be
removed by cutting and pasting methods. In particular, if one
interprets these solutions as a phantom scalar field collapse, the
singularities can be easily removed.

There seems to be much prejudice and suspicion against fields with
NKE, especially, due to their possible vacuum instability. The
stability is important, however, depending on which problem is to be
addressed. If we approach the cosmological singularity, the decay of
the fields with NKE should not pose conceptual difficulties.  One
needs these fields just for short times to smoothen the singularity,
and when this is achieved they may as well decay: the ghost appears
and then disappears, leaving the world ghost-free. It would be
interesting to see whether the phenomenological approach considered
here can be obtained from field/string theories, we leave this,
however, for future works.

\section{Acknowledgements}

We are grateful to Jos\'e Senovilla for helpful discussions and valuable
comments.  A.F. was supported by the University of the Basque Country
Grants 9/UPV00172.310-14456/2002, and The Spanish Science Ministry
Grant 1/CI-CYT 00172. 310-0018-12205/2000. S.J. acknowledges support
from the Basque Government research fellowship.

\end{document}